\documentclass[floatfix,%
 reprint,
 amsmath,amssymb,
 aps,
]{revtex4-2}

\usepackage{graphicx}
\usepackage{dcolumn}
\usepackage[colorinlistoftodos]{todonotes}
\usepackage{bm}
\usepackage{ulem}
\usepackage{xcolor} 

\usepackage{tabularx}
\usepackage{multirow}
\usepackage{appendix}
\usepackage{array}
\usepackage{booktabs}   
\usepackage{enumitem}   

\begin{document}

\preprint{APS/123-QED}

\title{Experimental Skills for Undergraduate Career Preparation in Quantum Information Science and Engineering}

\author{Shams El-Adawy\textsuperscript{1,2}}

\author{A.R. Pi\~na\textsuperscript{3}}

\author{Benjamin M. Zwickl\textsuperscript{3}}

\author{H. J. Lewandowski\textsuperscript{1,2}}

\affiliation{\textsuperscript{1}JILA, National Institute of Standards and Technology and the University of Colorado, Boulder, Colorado 80309, USA}
\affiliation{\textsuperscript{2}Department of Physics, University of Colorado, Boulder, Colorado 80309, USA }
\affiliation{\textsuperscript{3}School of Physics and Astronomy, Rochester Institute of Technology, Rochester, NY 14623, USA}

\date{\today}

\begin{abstract}
The growth of the Quantum Information Science and Engineering (QISE) industry has increased interest in how undergraduate programs prepare students for careers in this field. Prior research emphasizes the value of experiential learning as preparation for the quantum industry, but lacks specificity regarding the experimental skills needed for positions available to bachelor's degree graduates. In this study, we investigate the experimental skills associated with bachelor's-level quantum industry positions through 44 semi-structured interviews with quantum industry professionals.  Guided by the American Association of Physics Teachers recommendations for the undergraduate physics laboratory curriculum, we characterize the experimental skills associated with positions described as requiring bachelor's-level preparation and thematically synthesize them into four categories: instrumentation, computation and data analysis, experimental and project design, and communication and collaboration. We further examine how these skills cluster across role types and articulate them as learning goals to provide guidance for educators interested in aligning undergraduate instruction with the needs of students wanting to pursue a career in the quantum industry. Our findings suggest the need to emphasize the discussion of hardware in QISE theory courses, expand experimental training through instructional laboratories, and intentionally integrate professional skills in undergraduate QISE education. 
\end{abstract}

\maketitle


\section{Introduction}

Quantum Information Science and Engineering (QISE) is an interdisciplinary field rooted in physics that underpins the development of quantum technologies, including sensing, networking and communication, and computing. Over the past decade, QISE research and education have received increased national and international attention \cite{national_quantum_initiative_2018, riedel2019europes,knight_walmsley_2019, nstc_qist_workforce_plan_2022}, including the designation of 2025 as the International Year of Quantum Science and Technology \cite{unesco_international_year_quantum_science_technology_2025}. As the field evolves, questions about how undergraduate education can prepare students for participation in the quantum workforce have become increasingly salient.  

The QISE industry encompasses a broad range of technical roles that are not limited to PhD level preparation, with bachelor's-level professionals contributing to the design, development, and operation of quantum technologies \cite{fox2020preparing, greinert2024advancing, el2025industry}. These positions span hardware, software, business, public facing, and roles that bridge across technical areas \cite{pina2025categorization}. Across this breadth of work, many bachelor's-level positions involve hands-on experimental work, which motivates the question of which experimental skills are most relevant for bachelor's degree graduates entering the QISE workforce.

Addressing this question is challenging because of the differences between the type of work performed in industry by bachelor's degree graduates and current undergraduate QISE curricula. A majority of current QISE courses mainly focus on theoretical aspects of quantum information \cite{pina2025landscape, meyer2024introductory}, while workforce and education studies highlight that many QISE roles involve building, testing, integrating, and maintaining physical systems \cite{fox2020preparing, aiello2021achieving, asfaw2022building, greinert2023future, greinert2024advancing, QEDC_ConnectingTheDots_2025}. Although these studies call for increased experiential learning opportunities, they tend to describe these needs broadly, without specifying which experimental skills are needed for particular positions or how those skills can be translated into undergraduate learning goals. 

This lack of specificity is a barrier for educators wanting to revise existing courses or design new courses in response to quantum workforce needs \cite{el2025insights}. Educators lack information on  which skills to prioritize and how these priorities should be converted into course and program goals within their institutions' resources and constraints. Addressing this challenge requires moving beyond general calls for experiential learning toward a more detailed understanding of the experimental skills that define day-to-day work in quantum companies. 

In this paper, we address this need by characterizing the experimental skills used by bachelor’s level professionals in ways that can inform undergraduate curricula. By experimental skills, we refer to scientific skills involved in working with physical systems in laboratory environments. These skills include modeling \cite{zwickl2015model, 2018modelling}, designing experiments \cite{cai2021student}, troubleshooting experimental apparatus \cite{troubleshooting2016}, developing various technical and practical skills \cite{2014aaptlabrecommendations}, and communicating \cite{hoehn2020investigating, lewis2025surveying}.  While physics education research (PER) has examined how students develop these skills, the role of these skills' in preparing students for bachelor's-level QISE industry positions has not yet been investigated. Thus, this study addresses the following research questions:

\begin{itemize}
    \item[] \textbf{RQ1:} What positions in the quantum industry are available to bachelor's degree graduates?
    \item[] \textbf{RQ2:} What experimental skills are needed for positions available to bachelor’s degree graduates in the quantum industry?
    \item[]\textbf{RQ3:} How can these identified experimental skills inform learning goals for undergraduate courses?
\end{itemize}

Our goal with this paper is to provide a characterization of experimental skills in bachelor's-level quantum industry positions articulated as learning goals for actionable use by educators. We begin by situating our work within the broader literature on QISE workforce and education and physics education research (PER) on laboratory instruction in Sec. \ref{Background}. We then describe  our data collection and analysis methods in Sec. \ref{Methods}. Subsequently, in Sec. \ref{Results}, we  present the bachelor's-level positions identified in our dataset, and report the experimental skills articulated as learning goals associated with those positions. Lastly, in Sec. \ref{Discussion}, we synthesize our results and discuss implications for undergraduate education, followed by conclusions and directions for future work in Sec. \ref{Conclusion}.

\section{Background}\label{Background}
Understanding how experimental skills are used and valued in the quantum industry requires situating our research  at the intersection of the literature on quantum education and workforce development and PER on laboratory instruction. Both of these bodies of work provide complementary perspectives about the challenges and opportunities of preparing students for careers in QISE. Quantum education and workforce studies describe the rapidly evolving needs of the quantum industry and QISE in higher education, while PER on labs offers well-established literature on how experimental skills are conceptualized, taught, and assessed in undergraduate physics. Bringing these perspectives together allows us to articulate how experimental preparation in physics may support bachelor's degree holders entry into the QISE workforce. 

\subsection{Quantum education and workforce research}
Quantum workforce studies describe an industry that relies on all educational levels, including bachelor's degree graduates who can contribute to building, testing, operating, and maintaining experimental systems, while working in interdisciplinary teams and in a context characterized by rapid technological change \cite{fox2020preparing, hughes2022assessing, greinert2023future, greinert2024advancing, QEDC_ConnectingTheDots_2025}. Workforce projections further indicate a growing need for graduates trained at all educational levels and multiple disciplines, including bachelor's degree graduates in physics, to support the scaling and deployment of quantum technologies \cite{eladawy2025projectedworkforce}. 
These studies characterize the desired knowledge and skills in broad terms \cite{hughes2022assessing, greinert2024advancing, QEDC_ConnectingTheDots_2025}, which makes it challenging for educators  to translate workforce needs into specific learning goals.   

Research drawing on faculty and program directors' perspectives points to the need for greater clarity and specificity about which skills are most important for preparing undergraduates for QISE careers \cite{meyer2022today, el2025insights}.  This lack of clarity is especially pronounced for experimental work, which is valued by quantum industry, yet underspecified in the current literature \cite{fox2020preparing, greinert2024advancing, el2025industry}. From the student perspective, studies of their perceptions of quantum careers indicate that undergraduates are not always deeply aware of what the quantum industry is and does, and what kinds of work are typically performed by individuals with bachelor's degrees \cite{oliver2025education, oliver2025capstone, Ella2025QuantumInterest}.  In other words, students and faculty alike lack clarity about the specific skills that support participation in the quantum industry.

In parallel, quantum education researchers have examined how undergraduate programs are responding to the growth of the quantum industry. Analyses of the current quantum education landscape in the United States show a proliferation of quantum and QISE-related courses and programs \cite{meyer2024introductory, meyer2024disparities, pina2025landscape, kruse2025quantum, buzzell2025quantum}. This work has focused on mapping curricula offerings, identifying common instructional approaches, and characterizing emerging patterns across institutions. In particular, work on the QISE education landscape highlights that many QISE courses emphasize theoretical formalism and abstractions, with comparatively limited attention to experimental practice \cite{pina2025landscape}. 

Despite recent efforts to expand access to authentic quantum experimental experiences \cite{borish2023implementation, sun2025exploring, ivory2025qcamp}, there remain open questions about how experimental skills for the QISE workforce can be taught and developed. In particular, there is a need to further articulate how teaching experimental skills for the quantum context can be scaled across institutions, while fitting within the  practical constraints of undergraduate programs. Overall, the quantum workforce and education literature underscore the importance of experimental skills, while leaving under articulated what those skills entail and how they can be meaningfully integrated into the undergraduate curriculum. 

\subsection{Undergraduate physics laboratory research}

Research on undergraduate laboratory instruction provides a well-established perspective on the development of experimental skills. The literature in this space offers instructional frameworks and assessment approaches for understanding how student learn to engage in experimental practices within physics \cite{etkina2010design, 2014aaptlabrecommendations,wilcox2017developing, holmes2020developing, hoehn2020framework, walsh2022skills, borish2022modeling}. As such, PER on laboratories provides a useful foundation for examining experimental skills preparation in the context of quantum workforce needs. 

A widely used synthesis of early laboratory instruction is the American Association of Physics Teachers (AAPT) recommendations for the undergraduate physics laboratory curriculum developed in 2014 \cite{2014aaptlabrecommendations}. These recommendations position laboratories as spaces for developing authentic scientific practices, such as modeling physical systems, designing experiments, analyzing and visualizing data,  developing technical and practical laboratory skills, and communicating results \cite{2014aaptlabrecommendations}. In fact, these scientific practices define what is commonly meant by the phrase ``experimental skills'' within the context of undergraduate physics.

A substantial body of PER provides empirical evidence that laboratories emphasizing these skill-based goals can produce meaningful student learning outcomes. Studies of students' views about experimental physics show that  laboratories designed around skill development can support students in developing more expert-like views of experimental physics \cite{wilcox2017developing, wilcox2017students} and can improve students’ critical thinking skills  \cite{walsh2022skills}. These results highlight that skill-focused laboratory instruction represents one of the primary ways, beyond undergraduate research experiences, for supporting the development of lab-specific skills. 

Building on this evidence, subsequent research has examined the development of a subset of these laboratory-specific skills in greater depth. For example, modeling-focused studies investigate how features such as project goals and apparatus complexity in student-designed lab projects shape students' engagement in model construction and revisions \cite{borish2022modeling}. Complementary research on troubleshooting shows that students draw on modeling skills when they collaboratively troubleshoot  malfunctioning apparatus \cite{troubleshooting2016, van2017investigating}. This work on modeling showcases undergraduate physics lab courses as key spaces for engaging students in modeling as a core experimental practice. 

Additional research has examined students’ development of experimental design skills. Across various instructional contexts, studies show that providing students with choices in experimental decisions supports increased agency \cite{karelina2007acting, ownership2017student, holmes2020developing, cai2021student, kalender2021restructuring}. In these studies, researchers examine how opportunities for students to design their own experimental procedures within various instructional lab structures impact student learning and behavior \cite{wilcox2016open,liu2025students}. This research positions designing experiments as another core component of students' participation in authentic experimental practice.  

In parallel, research on writing in lab courses demonstrates that structured opportunities for written communication support students' understanding of scientific communication as a professional activity with specific associated norms and practices \cite{stanley2018recommendations, hoehn2020framework, hoehn2020investigating,hoehn2020incorporating}. In particular, written tasks in physics lab courses allow students to articulate experimental reasoning, interpret data, and coordinate collaborative work, which reinforces communication as an integral component of experimental practice rather than a peripheral skill.

Despite this extensive literature on the development of experimental skills in undergraduate physics labs to support engagement in authentic scientific practices, PER has largely focused on laboratory learning outcomes in relation to preparation for future coursework, academic research, or general STEM workforce. There has been comparatively little attention to how these skills align with the needs of a concrete and emerging employment context such as the quantum industry.  

Viewed holistically, while PER provides insight into the development of experimental skills and quantum workforce studies emphasize the value of experimental work, neither literature specifies how experimental skills manifest in bachelor's-level quantum industry positions. In this study, we address this limitation by examining experimental skills through the lens of quantum industry professionals. We use the AAPT recommendations \cite{2014aaptlabrecommendations}
as an initial disciplinary anchor,  while adapting them to reflect the emergent patterns from our data and interdisciplinary nature of QISE, as we discuss in more details in Sec. \ref{Data analysis}.

\section{Methods} \label{Methods}

\subsection{Data collection}

Our data consisted of research interviews with professionals working in quantum companies based in the United States. Recruitment consisted of both purposeful and convenience sampling \cite{etikan2016comparison}. We leveraged the research team's institutional networks, including alumni in the quantum industry, and used snowball sampling \cite{parker2019snowball} by asking each interviewee to refer colleagues within their company to participate in our research interviews. Concurrently, we asked the Quantum Economic Development Consortium (QED-C) \cite{QEDC_website}, which has more than 150 member companies, to solicit participation from professionals part of the consortium.  Across these recruitment strategies, responses rates varied. Individuals with existing professional or institutional connections to the research team were more likely to respond and participate in interviews, whereas outreach to companies or individuals without such connections often did not result in participation.

From December 2024 to January 2026, we conducted 44 semi-structured interviews over Zoom with employees and managers in quantum companies. Interviews lasted approximately one hour. To capture both high-level overview and task-level perspectives, we had two versions of the interview protocol, one for managers, and one for employees, which are included in the Supplemental Material. In brief, our protocols enabled managers to provide a broad view of different positions in their company and associated knowledge, skills, and abilities (KSAs), while employees focused on the tasks they personally engage in and their associated KSAs.

\begin{figure}
    \centering
    \includegraphics[width=1.0\linewidth]{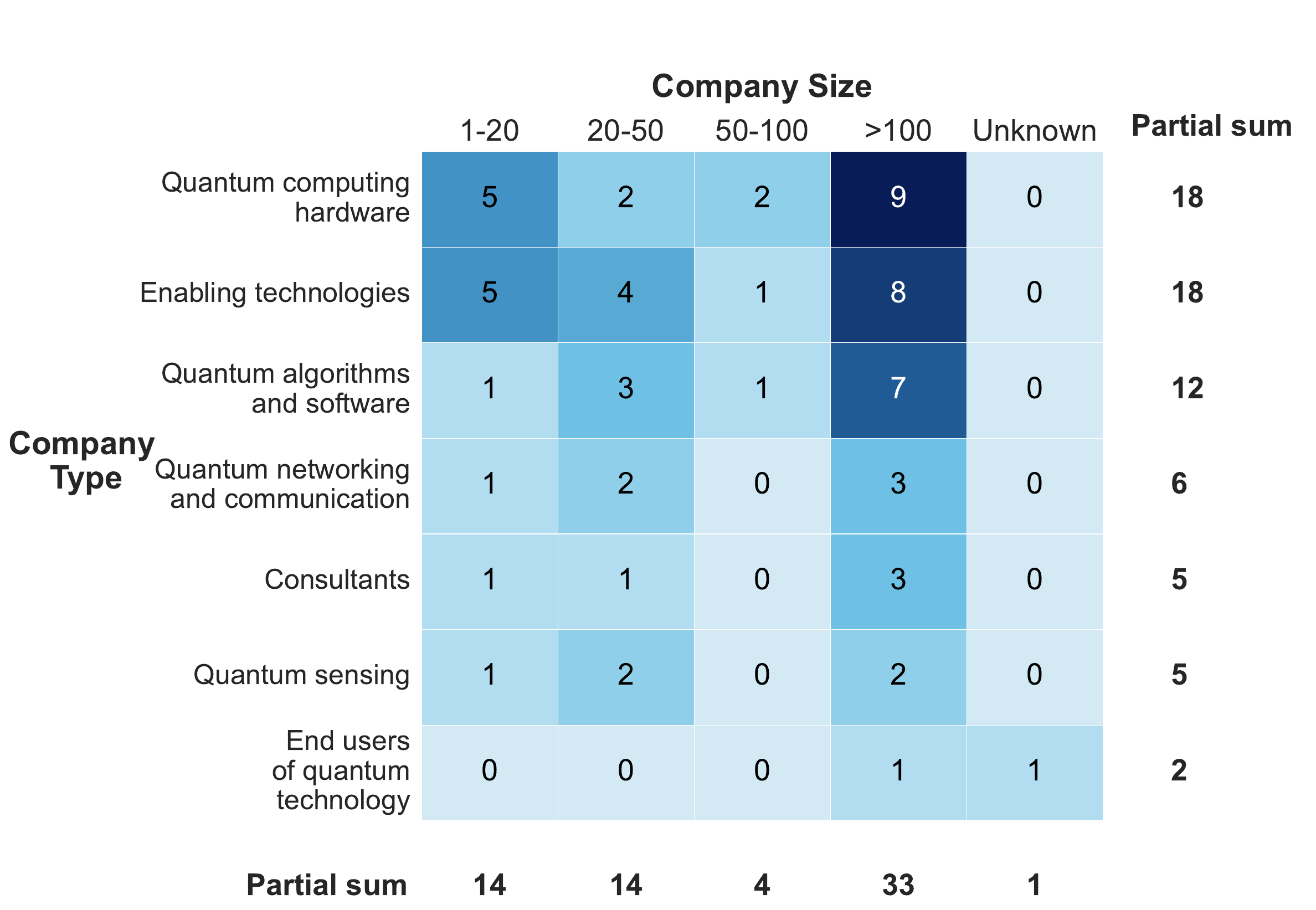}
    \caption{Distribution of the 24 distinct companies by type and size represented by the 44 interviews analyzed in this study. Companies are categorized by activity type following definitions in prior literature \cite{el2025industry}. Company size corresponds to the interviewee-reported range of employees working in quantum-related technologies in these companies. The numbers in each cell indicate the count of companies associated with a given company type (rows) and size category (columns). Because many companies report participation in multiple activity types, a single company may be counted in more than one row. Row-wise partial sums show the total number of companies for each company type across all sizes. Column-wise partial sums refer to the total number of companies within each size bin. As result, the sum of partial counts exceeds the number of distinct companies, since companies contributing to multiple activity types are counted multiple times.}
\label{fig:companycharacterization}
\end{figure}

Interviewees came from various company types and sizes (see Fig. \ref{fig:companycharacterization}), which allowed us to have insights into workforce needs from a variety of jobs in the quantum industry. In total, interviewees were affiliated with 24 distinct companies, several of which report engagement in multiple types of quantum-related activities. 

\subsection{Data analysis} \label{Data analysis}
Interviews were transcribed using professional human transcription \cite{gotranscript} and analyzed using thematic analysis \cite{braun2006using}. Initial codes captured company characteristics, position discussed in each interview, KSAs descriptions, tasks, educational level, and disciplinary background for each position. Consistent with our research questions, we focused on the tasks and KSAs relating to experimental work, rather than focus on quantum KSAs, which we explicitly address in separate work \cite{Pina2025QuantumWorkforce}.

Because managers described multiple positions within their company,  we had information about more positions than we had interviews.  Across the 44 interviews, we identified 88 distinct positions, and each position was treated as a distinct analytic unit. 

To align our analysis with our research questions, we then filtered the dataset to retain only positions that participants described as requiring bachelor's-level preparation. Positions described as requiring “Bachelor’s, Master’s, or PhD” were excluded, as the inclusion of PhD-level qualification implied openness to hiring PhD-level candidates, which was outside the scope of our research questions.  Positions described as requiring “Bachelor’s, Master’s” were included since their target qualification excluded PhDs. Although our filtering approach may under count positions in which bachelor's degree holders may be hired in practice, particularly for roles that may be open to multiple degree levels, it allowed us to center our analysis on positions where bachelor's degrees were clearly foregrounded in interviews. Our filtering process resulted in a final set of 24 positions.

Subsequently, to identify the experimental skills associated with these positions, we began with the AAPT recommendations for the undergraduate physics laboratory curriculum \cite{2014aaptlabrecommendations} as a disciplinary framework for our analysis. The AAPT recommendations provided a widely recognized articulation of experimental skills categories (e.g., designing experiments, modeling, communicating physics)  within undergraduate physics education \cite{holmes2019operationalizing}, helped operationalize what counted as experimental skills in our dataset and allowed us to map our findings directly onto instructional practice. Using the AAPT recommendations, we identified the KSAs discussing experimental skills in our data, which led to the creation of an initial list of experimental skills.

That initial list of skills highlighted that interviewees' descriptions of experimental skills embedded within workplace tasks often spanned multiple categories within the AAPT recommendations and covered multiple conceptual or procedural components within a single AAPT recommendations category. Given that many interviewees' responses blended across and within AAPT recommendations categories, imposing the original structure on our data risked losing information about how these skills actually appear in quantum industry roles. 
Thus, we removed the  original categories label, and decomposed skill definitions into smaller discrete units as described by interviewees.

Then, we engaged in an iterative refinement process to organize these skills into a coherent structure.  We examined the identified skills and coded existing interview segments to look for conceptual similarities and distinctions among the skills. Through successive discussions among the research team, skills were clustered into broader categories that reflected patterns in the data. This process resulted in four overarching categories of experimental skills:
(1) instrumentation, (2) computation and data analysis, (3) experimental and project design, and (4) communication and collaboration. Within each category, individual skills were articulated in learning goal language to provide specificity to educators wanting to concretely address those skills in their courses. 

Lastly,  following our prior work \cite{pina2025categorization}, we categorized each of the positions into one of four types based on similarities in tasks and KSAs, which reflects similarities in the use of experimental skills across positions. Our four types of categories were:
\begin{itemize}
    \item \textit{Hardware}, which refers to roles where individuals engage in hands-on work designing, developing, refining, and manufacturing hardware for quantum technologies.
    \item \textit{Software}, which refers to roles where individuals engage in the work of designing, developing, and optimizing software for quantum systems and applications.
    \item \textit{Bridging} which refers to roles where individuals focus on connecting teams within an organization (e.g., bridging the gap between technical applications and the underlying hardware or software).
    \item \textit{Public facing and business}, which refers to roles where individuals are connected by their shared responsibilities regarding leading the company and engaging with the public or clients.
\end{itemize}

A detailed description of the role categorization scheme is provided in \cite{pina2025categorization} and a summary figure of the categories is included in Appendix \ref{appendix:categoriesofroles}. These categories of roles were used to interpret patterns in skills, which we discuss in Sec. \ref{Patternsof experimental skills}.

To check the consistency of the coding, categorization, and skills definitions, we conducted inter-rater reliability (IRR) between the first and second authors of this paper. Treating all the quotes related to each position as a single distinct analytic unit, we independently characterized 20\% of the positions with associated experimental skills.  Before discussion, the two coders reached 85\% agreement when applying the experimental skills codes. Discrepancies were related to the language used in the skills definitions, which was refined, and led to reaching full agreement on this subset of the data after discussion. This coding consistency was sufficient given that our primary goal was to establish consistency in identifying whether a skill was mentioned for a given position anywhere in the interview, rather than a fine-grained count of how often a particular interviewee mentioned a particular skill since that was highly dependent on the level of detail provided by each interviewee. 

\subsection{Limitations}
Several limitations should be considered when interpreting our findings. First, although our data set consisted of 44 interviews, which is the largest quantum industry interview study to date, only a subset addressed bachelor's-level positions, which may not be representative of the full range of bachelor’s level opportunities  across the quantum industry. Second, the depth and specificity with which participants described tasks and associated KSAs varied, which may have limited our ability to capture all relevant experimental skills. Third, while the AAPT recommendations provided a productive starting point, our analytic approach involved adapting and reorganizing elements of the recommendations based on patterns in the data. Although this represents a  non-standard use of the recommendations,  we consider it appropriate given our goal of capturing experimental skills as they are described in the quantum industry, which, while grounded in physics, is inherently interdisciplinary. Despite our adaption, the recommendations  still shape the language of our results. 

\section{Results}\label{Results}
\subsection{Roles in the quantum industry for bachelor’s level graduates}

Table \ref{tab:inidvidualpositions}, which answers RQ1,  summarizes the individual positions identified in our dataset that participants described as requiring preparation at the bachelor's level. Each position is organized with the title of the individual position used by interviewees, the associated educational level, and the disciplinary background identified as relevant for being hired into the position. 

The positions in Table \ref{tab:inidvidualpositions} span multiple role types. Position 1-11 are hardware roles, position 12 is a bridging role, positions 13-21 are software roles, and positions 22-24 are public facing and business roles. This distribution highlights the range of employment types available to individuals with bachelor's degrees across the quantum workforce. Across these role categories, participants most frequently described physics, engineering, and computer science as relevant disciplinary backgrounds. Several positions were also associated with more than one disciplinary background, which reflects the interdisciplinary nature of QISE as described by interviewees. This pattern  is consistent with prior characterizations of QISE as an interdisciplinary field that offers courses and programs distributed across multiple academic departments \cite{pina2025landscape}.

 By making explicit the variety of bachelor's-level positions and role types present, Table \ref{tab:inidvidualpositions} provides necessary context for examining how experimental skills appear across different kinds of quantum roles.

\begin{table*}[htbp]
\caption{Titles of individual positions, education, and discipline for bachelor’s-level graduates in the quantum industry. The \textit{Type} column indicates role categories: H = hardware roles, B = bridging roles, S = software roles, P = public-facing or business roles. In the discipline column, \textit{other} denotes that no particular discipline was specified, and participants from a range of disciplines were considered.}
\label{tab:inidvidualpositions}
\begin{ruledtabular}
\begin{tabular}{clll}
Type &
Individual position title &
Education &
Discipline \\ \hline
H &
Lab technician &
Associate, Bachelor &
Engineering, Other \\

H &
Fabrication engineer &
Bachelor &
Math, Physics \\

H &
Photonics assembly technician &
Bachelor &
Engineering, Physics \\

H &
Assembly technician &
Bachelor &
Physics \\

H &
Quantum R\&D engineer &
Bachelor &
Engineering Physics \\

H &
Laser and optics engineer &
Bachelor &
Chemistry, Physics \\

H &
Construction specialist &
Bachelor &
Other \\

H &
Junior photonics experimenter &
Bachelor &
Engineering (Optical), Physics \\

H &
Research engineer &
Bachelor, Master &
Chemistry, Computer Science, Engineering, Physics \\

H &
Research scientist / quantum systems engineer &
Bachelor, Master &
Engineering (Thermal and Mechanical), Physics \\

H &
Quantum engineer &
Bachelor, Master &
Engineering, Engineering Physics \\

B &
System operator &
Bachelor &
Engineering, Physics \\

S &
Quantum software developer 1 &
Bachelor &
Computer Engineering \\

S &
Quantum software developer 2 &
Bachelor &
Computer Science \\

S &
Quantum software engineer 1 &
Bachelor &
Computer Science, Engineering, Math \\

S &
Quantum software engineer 2 &
Bachelor &
Computer Science, Math, Physics \\

S &
Scientific software developer &
Bachelor &
Computer Science \\

S &
Firmware developer &
Bachelor &
Engineering (Electrical) \\

S &
Algorithm developer 1 &
Bachelor, Master &
Computer Science, Physics, \\

S &
Algorithm developer 2 &
Bachelor, Master &
Computer Science, Math, Physics, Other \\

S &
Junior scientific software engineer &
Bachelor, Master &
Computer Science \\

P &
Government industry advocate &
Bachelor &
Applied Physics \\

P &
Education advocate &
Bachelor, Master &
Computer Science, Math, Physics \\

P &
Chief Operating Officer &
Bachelor &
Engineering, Other \\
\end{tabular}
\end{ruledtabular}
\end{table*}

\subsection{Experimental skills as learning goals for bachelor’s level graduates}\label{Experimental skills for bachelor’s level graduates}
Building on the characterization of bachelor's-level quantum industry positions presented in Table \ref{tab:inidvidualpositions}, we present here the experimental skills that interviewees associated with these positions. Across the dataset, interviewees describe experimental work as encompassing a broad range of hands-on collaborative tasks and activities. As described in Section \ref{Data analysis}, we organize these skills into four categories: instrumentation, computation and data analysis, experimental and project design, and communication and collaboration. These categories reflect the integrated and cross-functional nature of experimental work in QISE positions and underscore the extent to which hands-on experimental contributions form a core component of these positions. Throughout this section, we include  representative interview excerpts and task descriptions to illustrate how skills manifested in practice for the quantum professionals in our dataset.  The primary result (our answers to RQ2 and RQ3) is four tables of learning goals (Tables \ref{tab:instrumentation_skills}, \ref{tab:computation_data_skills}, \ref{tab:experimental_project_skills}, and \ref{tab:communication_collaboration_skills})that follow each category's results, which synthesize the experimental skills as set of learning goals for educators. 

\subsubsection{Instrumentation}

Instrumentation refers to operating, maintaining, and troubleshooting scientific and engineering instruments and systems. Interviewees emphasize that bachelor’s level employees are often working directly with hardware, especially in tasks involving optics, electronics, and vacuum systems. 
For example, Alex, who works at a company developing quantum computing hardware, algorithms, and software, highlights the value placed on prior optics experience when hiring for a \textit{laser and optics engineer} position. According to Alex, candidates who had worked with optical systems during their undergraduate education are often better prepared for this position:
\begin{quote}
    I would just say optics experience. A lot of our employees did have a bachelor's in physics, but they worked in an undergrad atomic physics lab or something, doing some optics, or maybe they were similarly in chemistry doing some optics undergrad chemistry experience. That is often a differentiator when hiring for this role.
\end{quote}
This emphasis on prior optics experience illustrates how familiarity with optical systems, which could be developed during undergraduate research laboratory experiences, is valued in the quantum industry. Alex further explains how familiarity with basic electronic measurement tools and experience with interferometers are also foundational instrumentation skills for this bachelor's-level employee:
\begin{quote}
    Understanding how a multimeter works and what is impedance and these basic things are helpful. We also do interferometry measurements. We work with optical surfaces, so we need to measure their flatness using an interferometer.
\end{quote}

According to Alex, common tasks for this \textit{laser and optics engineer} include: assisting in running optics and photonics experiments, designing and maintaining optical systems, defining hardware specifications and procuring necessary equipment, and setting up lasers and beam lines.

These examples of instrumentation skills and tasks illustrate that instrumentation at the bachelor's-level is not limited to routine operation of equipment, but also encompass contributing to the design and iterative refinement of experimental setups.  

Table \ref{tab:instrumentation_skills} summarizes the instrumentation skills articulated as learning goals identified across our dataset, which range from operating common laboratory equipment and working with optical and electronic systems to troubleshooting experimental setups.

\begin{table*}[htbp]
\caption{Learning goals for instrumentation skills (I1--I9) for quantum industry positions held by bachelor's graduates.}
\label{tab:instrumentation_skills}
\begin{ruledtabular}
\begin{tabular}{ll}
I1 & \begin{minipage}[t]{0.85\textwidth}\raggedright
Able to operate basic test and measurement devices (e.g., power supplies, multimeters)
\end{minipage} \\
I2 & \begin{minipage}[t]{0.85\textwidth}\raggedright
Able to work with electronic systems and components
\end{minipage} \\
I3 & \begin{minipage}[t]{0.85\textwidth}\raggedright
Able to operate and fix laser systems
\end{minipage} \\
I4 & \begin{minipage}[t]{0.85\textwidth}\raggedright
Able to work with optical systems and perform optical alignment
\end{minipage} \\
I5 & \begin{minipage}[t]{0.85\textwidth}\raggedright
Able to operate interferometers to assess surface flatness
\end{minipage} \\
I6 & \begin{minipage}[t]{0.85\textwidth}\raggedright
Able to work with ultra-low-temperature apparatus
\end{minipage} \\
I7 & \begin{minipage}[t]{0.85\textwidth}\raggedright
Able to work with vacuum systems
\end{minipage} \\
I8 & \begin{minipage}[t]{0.85\textwidth}\raggedright
Able to do basic troubleshooting using an iterative and logical approach
\end{minipage} \\
I9 & \begin{minipage}[t]{0.85\textwidth}\raggedright
Develop general practical experimental skills through research experiences or lab coursework
\end{minipage} \\
\end{tabular}
\end{ruledtabular}
\end{table*}

\subsubsection{Computation and data analysis}
Computation and data analysis refer to applying computational tools to acquire, process, and interpret data. Interviewees describe computation as tightly integrated with experimental work, particularly for controlling instruments, interacting with hardware components, and data collection and analysis. Rather than being treated as a separate or purely theoretical activity, computational work is often embedded within day-to-day experimental tasks. For instance, Avery, who works at an enabling technologies and quantum computing hardware company, describes expectations for a \textit{photonics assembly technician} to be:
\begin{quote}
 Lab skills such as being able to work with optics in the lab, able to use laboratory electronics, and able to do some basic coding and data analysis.
\end{quote}
This \textit{photonics assembly technician} illustrates how coding and data analysis are treated as core experimental skills alongside optics and lab electronics.

Similarly, Jamie, who works at a company that does quantum networking and communication, sensing, algorithms, software, and enabling technologies, emphasizes programming skills for data analysis as essential for an entry level \textit{photonics experimenter} who would use:
\begin{quote}
That general scientific programming language [(e.g., Python)] to do the analysis and statistics.
\end{quote}
In particular, this \textit{photonics experimenter} would use statistics and data analysis to reduce the data and to be able to identify problems and trends in quantum experiments.

These examples of skills and tasks showcase that computation \& data analysis skills are valued for their practical utility in supporting experimental setups, interpreting data, and guiding iterative refinement. Table \ref{tab:computation_data_skills} summarizes the computation and data analysis skills identified in our dataset, which include having general programming proficiency, selecting appropriate computational tools, troubleshooting code, and performing data analysis, including uncertainty analysis. 

\begin{table*}[htbp]
\caption{Learning goals for computation and data analysis skills (D1--D6) for quantum industry positions held by bachelor's graduates.}
\label{tab:computation_data_skills}
\begin{ruledtabular}
\begin{tabular}{ll}
D1 & \begin{minipage}[t]{0.85\textwidth}\raggedright
Able to interpret computational or measurement results to understand physical systems
\end{minipage} \\
D2 & \begin{minipage}[t]{0.85\textwidth}\raggedright
Able to do some programming in general, without specifying whether it is for data processing, data representation, or other purposes
\end{minipage} \\
D3 & \begin{minipage}[t]{0.85\textwidth}\raggedright
Able to select numerical algorithms or software tools based on problem requirements
\end{minipage} \\
D4 & \begin{minipage}[t]{0.85\textwidth}\raggedright
Able to troubleshoot code
\end{minipage} \\
D5 & \begin{minipage}[t]{0.85\textwidth}\raggedright
Able to use computers to collect and process data
\end{minipage} \\
D6 & \begin{minipage}[t]{0.85\textwidth}\raggedright
Able to perform data analysis, including uncertainty analysis
\end{minipage} \\
\end{tabular}
\end{ruledtabular}
\end{table*}

\subsubsection{Experimental and project design}
Experimental and project design refers to planning, executing, and managing experiments. Interviewees highlight how bachelor’s level employees contribute to designing and refining experiments by thinking through requirements, constraints, and tradeoffs for technologies that are still emerging and procedures that are not yet standardized. 

For example, Casey, who works at a quantum computing hardware company, describes expectations for a \textit{quantum engineer} to have a working understanding of basic systems engineering practices to define hardware requirements for different systems:
\begin{quote}
    Yes, I'd say a little bit of systems engineering, understanding requirements, when good is good enough.
\end{quote}
This interviewee highlights the importance of understanding requirements and determining acceptable performance thresholds to develop a device for a particular application. These skills would also support this \textit{quantum engineer} in integrating subsystems in a device, testing and calibrating hardware systems, as well as working on creating a product out of experimental results. 

Moreover, Avery, who works at an enabling technologies and quantum computing hardware company, highlights another aspect of designing experiments when describing the position of a  \textit{construction specialist} who is responsible for supporting quantum experiments through facility design and building:
\begin{quote}
   For this position, the specific skill is related to constructing these facilities somewhat specific to quantum: a cleanroom facility, a chip fabrication facility. [...] Being able to read and draw architectural plans in how buildings and cleanrooms are constructed.
\end{quote}
These illustrative tasks and skills show how experimental and project design skills for bachelor's-level graduates span both laboratory-scale experiments and larger infrastructure considerations. Table \ref{tab:experimental_project_skills} summarizes the experimental and project design skills that were identified across our dataset. These skills include planning and executing experiments, integrating multiple scientific principles, managing technical projects, and persisting through experimental setbacks. 
\begin{table*}[htbp]
\caption{Learning goals for experimental and project design skills (E1--E10) for quantum industry positions held by bachelor's graduates.}
\label{tab:experimental_project_skills}
\begin{ruledtabular}
\begin{tabular}{ll}
E1  & \begin{minipage}[t]{0.85\textwidth}\raggedright
Able to integrate multiple scientific principles when designing experiments
\end{minipage} \\
E2  & \begin{minipage}[t]{0.85\textwidth}\raggedright
Able to integrate multiple scientific principles to translate findings into practical applications
\end{minipage} \\
E3  & \begin{minipage}[t]{0.85\textwidth}\raggedright
Able to design a procedure to test a model or hypothesis
\end{minipage} \\
E4  & \begin{minipage}[t]{0.85\textwidth}\raggedright
Able to effectively plan and carry out experiments
\end{minipage} \\
E5  & \begin{minipage}[t]{0.85\textwidth}\raggedright
Able to manage complex technical projects from formulation through testing to reporting
\end{minipage} \\
E6  & \begin{minipage}[t]{0.85\textwidth}\raggedright
Able to perform basic systems engineering tasks, including understanding requirements and determining acceptable performance thresholds
\end{minipage} \\
E7  & \begin{minipage}[t]{0.85\textwidth}\raggedright
Able to design and oversee construction of specialized scientific facilities, including cleanrooms and chip fabrication facilities for quantum research
\end{minipage} \\
E8  & \begin{minipage}[t]{0.85\textwidth}\raggedright
Able to critically read scientific literature to refine a research question or optimize an experimental design
\end{minipage} \\
E9  & \begin{minipage}[t]{0.85\textwidth}\raggedright
Able to conduct an experiment in an open-ended and ill-defined context (e.g., working with ambiguity of emerging technologies)
\end{minipage} \\
E10 & \begin{minipage}[t]{0.85\textwidth}\raggedright
Able to persist through experimental setbacks, including adapting approaches until reliable outcomes are achieved
\end{minipage} \\
\end{tabular}
\end{ruledtabular}
\end{table*}

\subsubsection{Communication and collaboration}
Communication and collaboration refer to sharing technical insights, coordinating work, and engaging with multiple stakeholders. Interviewees consistently highlight that experimental work in the quantum industry is highly collaborative, which requires bachelor's-level employees to work effectively with colleagues with various technical expertise.

Logan, who works at an enabling technologies company, articulates the centrality of teamwork and communication for a \textit{research engineer}:
\begin{quote}
    Communication and Problem solving. Every problem is you're working on a team. You need folks that work with the software folks, work with the electrical engineers, work with the mechanical engineers, work with the manufacturing engineers. I just can't stress it enough. It's got to be solution-driven, team-focused. The goal is we're getting a product released to production.
\end{quote}
This \textit{research engineer} would need to communicate and problem solve with their team as they design and fabricate classical circuits, design and fabricate jigs, as well as design and fabricate mechanical housings. The interplay between these skills and tasks shows how communication is inseparable from  experimental progress, rather than a secondary professional skill. 

Furthermore, Morgan, who works at a company that does quantum computing hardware, algorithms, and software, underscores the importance of communication and collaboration for a  \textit{government-industry advocate}, especially when engaging with nontechnical stakeholders:
\begin{quote}
    Working with local state representative to communicate how to best help implement a quantum bill in the state.  I had to reach out to the representative, and then try to schedule a meeting, and communicate how we can best help. [...] If they're looking for a better understanding of how the industry works, as well as how the technology works, they need us to communicate that. 
\end{quote}
This example as well as the previous one  showcase that communication and collaboration skills support both internal coordination of experimental work and external engagement. Table \ref{tab:communication_collaboration_skills} summarizes the communication and collaboration skills identified across our dataset, which include  preparing presentations, maintaining technical documentation, writing reports, and collaborating effectively within teams. 

\begin{table*}[htbp]
\caption{Learning goals for communication and collaboration skills (C1--C5) for quantum industry positions held by bachelor's graduates.}
\label{tab:communication_collaboration_skills}
\begin{ruledtabular}
\begin{tabular}{ll}
C1 & \begin{minipage}[t]{0.85\textwidth}\raggedright
Able to prepare and deliver presentations to multiple stakeholders
\end{minipage} \\
C2 & \begin{minipage}[t]{0.85\textwidth}\raggedright
Able to collaborate effectively with others
\end{minipage} \\
C3 & \begin{minipage}[t]{0.85\textwidth}\raggedright
Able to create and maintain clear and up-to-date technical documentation for experimental setups and procedures
\end{minipage} \\
C4 & \begin{minipage}[t]{0.85\textwidth}\raggedright
Able to write reports and proposals
\end{minipage} \\
C5 & \begin{minipage}[t]{0.85\textwidth}\raggedright
Able to think critically (e.g., critical assessment of their own work and that of others to identify strengths and limitations)
\end{minipage} \\
\end{tabular}
\end{ruledtabular}
\end{table*}

The experimental skills identified in this section are reported as learning-goals in Tables \ref{tab:instrumentation_skills}, \ref{tab:computation_data_skills}, \ref{tab:experimental_project_skills}, and \ref{tab:communication_collaboration_skills} to support instructional use. In this way, the tables of learning goals provide a compact, transferable description of experimentally relevant skills for bachelor's-level QISE industry work that can be adapted as course level outcomes.

\subsection{Patterns of experimental skills among quantum industry positions available to bachelor’s level graduates}\label{Patternsof experimental skills}

\begin{figure*}
    \centering
    \includegraphics[width=0.6\linewidth]{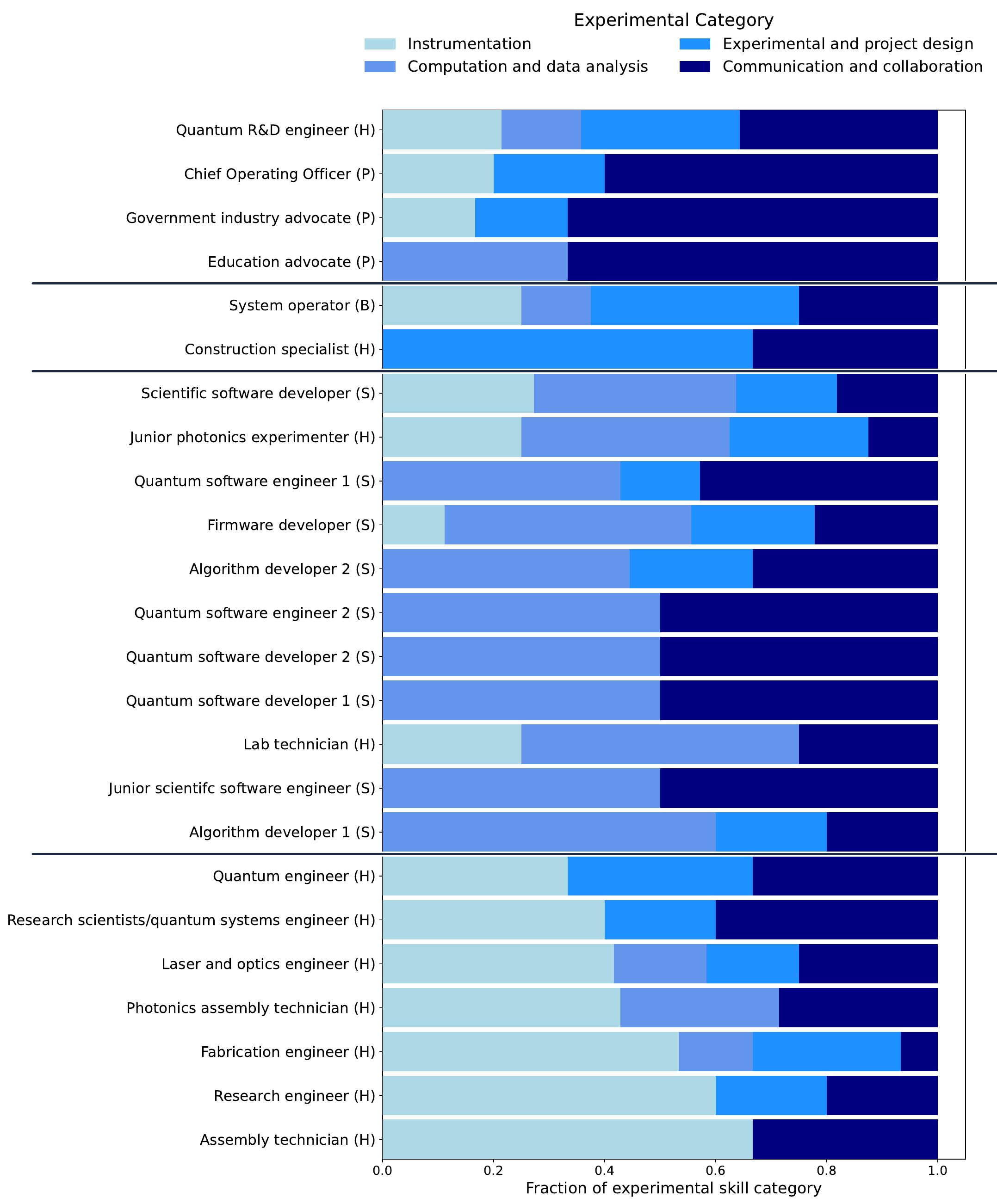}
    \caption{Distribution of experimental skills for each position grouped by dominant experimental skills category. The individual positions serving as starting and endpoint points for each category are: Quantum engineer → assembly technician emphasize instrumentation skills. Scientific software developer →  algorithm developer 1 emphasize computation and data analysis skills. System operator → construction specialist emphasize experimental and project design skills. Quantum R\&D engineer → education advocate emphasize communication and collaboration skills. Labels appended to each position indicate the role category type: hardware (H), software (S), bridging (B), and public facing and business (P) roles.}
    \label{fig:patternexperimentalskills}
\end{figure*}

To examine how experimental skills are distributed across bachelor's-level quantum industry positions, we situate each bachelor-level position within the four role categories introduced earlier (hardware, software, bridging, and public facing and businesses), and map the associated experimental skills identified. Table \ref{tab:nonphd_jobs} provides an overview of which experimental skills appear in which individual position, and Fig. \ref{fig:patternexperimentalskills} summarizes the relative prominence of these skills across positions.

Several predictable and consistent patterns emerge when positions are grouped by role type. Hardware-oriented positions heavily emphasize instrumentation skills. These positions (e.g., fabrication engineer, quantum engineer) often combine instrumentation skills with computation \& data analysis, particularly for operating equipment, monitoring system performance, and interpreting experimental data. In contrast, software-oriented roles (e.g., quantum software developer, quantum software engineer) draw most strongly on computation \& data analysis, with comparatively limited emphasis on instrumentation skills. While instrumentation skills do appear in software roles they are typically associated with interfacing software with experimental systems or supporting software-hardware workflows. 

Bridging roles exhibit the broadest range of experimental skills across categories. The one position classified as a bridging role (e.g., system operator) draws on experimental \& project design, communication \& collaboration, computation \& data analysis, and instrumentation. This distribution reflects the integrative nature of this role, which needs fluency across multiple experimental skill categories. Public facing and business oriented roles (e.g., education advocate, government industry advocate) rely the most on communication and collaboration, while still needing some familiarity with the other experimental skills categories to effectively engage with technical stakeholders. 

Across all role types, communication and collaboration is the only experimental skill category that appears in every position in our dataset. As shown in Table \ref{tab:nonphd_jobs}, all bachelor's-level positions include at least one skill related to communication and collaboration. This finding underscores the importance of collaborative work in quantum industry contexts and highlights communication as a foundational component of experimental practice in the quantum workforce. 

In summary, these patterns indicate that bachelor's-level positions cluster in systematic ways across the four experimental skills categories with distributions that align closely with broader role types.  Figure 
\ref{fig:patternexperimentalskills} visualizes these trends by showing the relative contribution of each experimental skill category across positions, which provides a consolidated view of how experimental work is structured across the bachelor's-level quantum workforce.

\begin{table*}[t]
\caption{Overview of the 24 individual positions and mapping to the four experimental skill categories. The skill codes reference specific skills in Tables~\ref{tab:instrumentation_skills}, \ref{tab:computation_data_skills}, \ref{tab:experimental_project_skills}, and \ref{tab:communication_collaboration_skills}.}
\label{tab:nonphd_jobs}
\begin{ruledtabular}
\begin{tabular}{lcccc}

\begin{minipage}[t]{0.22\textwidth}\raggedright
\textbf{Title of individual position}
\end{minipage}
&
\begin{minipage}[t]{0.18\textwidth}\raggedright
\textbf{Instrumentation}
\end{minipage}
&
\begin{minipage}[t]{0.18\textwidth}\raggedright
\textbf{Computation \& data analysis}
\end{minipage}
&
\begin{minipage}[t]{0.18\textwidth}\raggedright
\textbf{Experimental \& project design}
\end{minipage}
&
\begin{minipage}[t]{0.18\textwidth}\raggedright
\textbf{Communication \& collaboration}
\end{minipage}
\\ \hline

\begin{minipage}[t]{0.26\textwidth}\raggedright Lab technician\end{minipage}
& I9 & D2, D5 &  & C2 \\

\begin{minipage}[t]{0.26\textwidth}\raggedright Fabrication engineer\end{minipage}
& I1--I8 & D1, D5 & E1, E2, E5, E10 & C2 \\

\begin{minipage}[t]{0.26\textwidth}\raggedright Photonics assembly technician\end{minipage}
& I2, I4, I9 & D2, D6 &  & C1, C5 \\

\begin{minipage}[t]{0.26\textwidth}\raggedright Assembly technician\end{minipage}
& I2, I3, I4, I9 &  & & C1, C2 \\

\begin{minipage}[t]{0.26\textwidth}\raggedright Quantum R\&D engineer\end{minipage}
& I2, I4, I8 & D2, D6 & E3, E4, E6, E8 & C1--C5 \\

\begin{minipage}[t]{0.26\textwidth}\raggedright Laser and optics engineer\end{minipage}
& I3, I4, I5, I8, I9 & D2, D3 & E6, E9  & C1, C2, C5  \\

\begin{minipage}[t]{0.26\textwidth}\raggedright Construction specialist\end{minipage}
&  &  & E5, E7 & C1 \\

\begin{minipage}[t]{0.26\textwidth}\raggedright Junior photonics experimenter \end{minipage}
& I2, I4  & D1, D2, D6 & E1, E4 & C4 \\

\begin{minipage}[t]{0.26\textwidth}\raggedright Research engineer\end{minipage}
& I3, I4, I8 &  & E5 & C2 \\

\begin{minipage}[t]{0.26\textwidth}\raggedright Research scientist / quantum systems engineer\end{minipage}
& I8, I9 &  & E10 & C2, C5 \\

\begin{minipage}[t]{0.26\textwidth}\raggedright Quantum engineer\end{minipage}
& I3 &  & E6 & C2 \\
\begin{minipage}[t]{0.26\textwidth}\raggedright System operator\end{minipage}
& I1, I8 & D1 & E1, E2, E6 & C1, C3 \\

\begin{minipage}[t]{0.26\textwidth}\raggedright Quantum software developer 1\end{minipage}
&  & D1, D2, D3 & & C1, C2, C3 \\

\begin{minipage}[t]{0.26\textwidth}\raggedright Quantum software developer 2\end{minipage}
&  & D1, D2, D4 &  & C1, C2, C5 \\

\begin{minipage}[t]{0.26\textwidth}\raggedright Quantum software engineer 1\end{minipage}
&  & D1, D2, D4 & E10 & C1, C2, C5 \\

\begin{minipage}[t]{0.26\textwidth}\raggedright Quantum software engineer 2\end{minipage}
&  & D1, D2 &  & C2, C5 \\

\begin{minipage}[t]{0.26\textwidth}\raggedright Scientific software developer \end{minipage}
&  I1, I2, I8 & D1, D2, D3, D4 & E2, E6  & C1, C2 \\
\begin{minipage}[t]{0.26\textwidth}\raggedright Firmware developer \end{minipage}
&  I2 & D2, D3, D4, D6 & E3,E5 & C2, C5 \\
\begin{minipage}[t]{0.26\textwidth}\raggedright Algorithm developer 1\end{minipage}
&  & D3, D4, D6 & E3 &  C5 \\

\begin{minipage}[t]{0.26\textwidth}\raggedright Algorithm developer 2\end{minipage}
&  & D1, D2, D3, D4 & E8, E10 & C1, C2, C5 \\

\begin{minipage}[t]{0.26\textwidth}\raggedright Junior scientific software engineer\end{minipage}
&  & D2, D3 &  & C3, C5 \\

\begin{minipage}[t]{0.26\textwidth}\raggedright Government industry advocate\end{minipage}
& I4 &  & E6 & C1, C2, C4, C5 \\

\begin{minipage}[t]{0.26\textwidth}\raggedright Education advocate\end{minipage}
& &  D2 &  & C1,C2 \\

\begin{minipage}[t]{0.26\textwidth}\raggedright Chief Operating Officer\end{minipage}
& I8 &  & E5 & C1, C2, C4 \\

\end{tabular}
\end{ruledtabular}
\end{table*}

\section{Discussion}\label{Discussion}

\subsection{Synthesis}
In this study, we characterize the experimental skills associated with quantum industry positions available to bachelor's-level graduates and situate those skills within broader role types across the QISE workforce. Building on prior quantum workforce and education studies that highlight the value of experimental work in the quantum industry \cite{fox2020preparing, hughes2022assessing, greinert2024advancing, QEDC_ConnectingTheDots_2025}, our findings provide a more concrete description of experimental skills as they appear in bachelor's-level quantum positions and how they cluster across different role types.

Across the positions in our dataset, experimental work was described as inherently cross-functional, integrating hands-on interaction with physical systems, computational tools, experimental project design, and collaboration within interdisciplinary teams. The four categories of experimental skills reflect how interviewees articulated experimental practice in their company context. While the depth and specificity of these descriptions varied across interviews, the consistency with which these categories appeared across positions suggests that they capture salient dimensions of experimental work for bachelor's-level positions in the quantum industry. 

As shown in Sec. \ref{Patternsof experimental skills}, experimental skills cluster in systematic ways that align with broader role types. Hardware roles emphasize instrumentation skills, software roles emphasize computation and data analysis, bridging roles draw on all four categories, and public facing and business roles rely most heavily on communication and collaboration. These patterns reflect the dominant emphases described by participants, not mutually exclusive categories of useful skills. In fact, many positions draw on multiple experimental skill categories, which underscores that the quantum industry values cross-functional professionals rather than narrowly specialized skill sets at the bachelor's-level.  

Notably, communication and collaboration emerges as the only experimental skill category present across all positions. Although interviewees varied in how explicitly they describe communication and collaboration tasks, the presence of this subset of skills across roles highlights that these professional skills are not just peripheral to experimental work in the quantum industry, but an integral component of it. This finding reinforces prior studies that emphasize the importance of communication, teamwork, and collaboration in quantum industry settings \cite{oliver2025capstone, oliver2025education}, though these skills are also important more broadly for STEM and non-STEM employers \cite{hoehn2020framework, lewis2025surveying,AACU2025Agility}.

Overall, our results extend existing quantum workforce and education studies by moving beyond broad calls for experiential learning to articulate how experimental skills are applied in bachelor's-level quantum industry positions. By connecting quantum industry professionals descriptions of experimental practice with learning goal language familiar to educators, our study bridges between research on the quantum workforce and PER on laboratory instruction to provide actionable implications for QISE education.

\subsection{Implications}
The experimental skills learning goals presented in Sec. \ref{Results} have several implications for undergraduate QISE education. Rather than solely creating entirely new curricular offerings, our findings point to opportunities for reorienting existing courses to more explicitly reflect the kinds of experimental work performed in the quantum industry.

First, undergraduate QISE courses can more explicitly situate theoretical content within experimental contexts. While many current QISE courses emphasize abstract concepts (e.g., qubits and gates without reference to hardware) \cite{pina2025landscape}, our results show various ways bachelor's-level employees engage with experiments. Explicitly integrating discussion of how theoretical concepts connect to system behavior, measurement, noise, and potential performance tradeoffs can help better prepare undergraduate students in understanding how theory informs experimental work in the quantum industry. Interestingly, this integration can be done even in courses without formal lab components. For example, learning goals such as E1 (integrating multiple scientific principles when designing experiments) and E2 (translating findings into practical applications) can be addressed through structured activities in which students analyze realistic device constraints, noise sources, and performance tradeoffs using the theory introduced in class. Similarly, E8 (critically reading scientific literature) can be addressed within QISE theory courses through guided analysis of experimental papers in which students evaluate methods, identify limitations, and propose refinements of experimental designs.  

Second, skills-focused instructional laboratories emerge as a particularly important environment for developing experimental skills relevant to the quantum workforce.  Many of the skills characterized in our data overlap with experimental skills already emphasized in undergraduate physics laboratories and are transferable across a range of career paths, which make skills-focused lab courses key in preparing students for QISE careers. 

For instance, instrumentation (I1-I8) and computation \& data analysis learning goals (D1-D6) are particularly well aligned with common objectives of skills-focused labs in first year and beyond-first-year bachelor's-level courses where students build their own apparatus, choose their own analysis method, and troubleshoot problems with the setup or apparatus \cite{zwickl2013process, holmes2020investigating}. In addition, learning goals associated with experimental design and project design such as E4 (planning and carrying out experiments) and E10 (persisting through setbacks) are particularly foregrounded in multi-week project-based labs where students engage in multiple scientific practices and  iteration \cite{werth2022impacts, werth2023enhancing, MerrittLewandowski2024, KretchmerMerrittLewandowski2024, Merritt2025StudentPerspectives}. 

Notably, E9 (conduct an experiment in an open-ended and ill-defined context) aligns with ongoing efforts from physics educators to incorporate open-ended activities in lab courses rather than simply following prescriptive laboratory guides to engage  students in more authentic experimental practices \cite{wilcox2016open, hoehn2021remote, liu2025students}. Furthermore, given that computation \& data analysis skills (D1-D6) primarily support experimental workflows in our dataset, integrating computational skills more explicitly in physics lab courses, an approach already adopted by some educators \cite{tufino2025using, Chap6IntegratingComputing, Chap7IntegratingComputing}, provides an additional opportunity for contextualizing these skills in support of quantum industry  career preparation.  

Third, the presence of communication and collaboration skills across bachelor's-level quantum industry roles highlights the need to continue to treat these practices as integral components of experimental skills rather than as peripheral professional skills. For example, labs that require students to maintain shared technical documentation, communicate experimental progress and results to peers and instructors,  coordinate task division and decision making within teams, and critically engage with experiments (C1-C5) reflect how experimental work is carried out in the quantum industry. While some existing lab courses already include communication, teamwork, collaboration, and critical thinking as explicit learning goals \cite{foote2018implementing, hoehn2020investigating,smith2020criticalthinking, werth2022assessing}, our results provide a way to contextualize such practices within an authentic employment context, thereby strengthening the relevance of these skills even more by highlighting their importance for positions in the quantum industry.

Finally, the distribution of skills across role types suggests that emphasizing broadly applicable experimental skills provides a scalable approach that can accommodate many variations in faculty expertise and institutional priorities
\cite{el2025insights}. Educators can adopt the learning goals that would most align with their instructional context, resources, and constraints. Importantly, the suggested instructional implications do not require complete curricular overhaul or the purchase of very expensive specialized apparatus, unlike approaches that rely on hands-on access to costly, specialized quantum hardware that may not be feasible across institutions with differing resources and expertise. In fact, the implications illustrate how experimental skills, already present in some undergraduate courses, can be made more explicit, intentional, and connected to the kinds of experimental work graduates are likely to encounter in the quantum industry. 

\section{Conclusion and Future work}\label{Conclusion}

Our study characterizes experimental skills associated with quantum industry positions available to bachelor’s  graduates and articulates those skills as learning goals. By mapping experimental skills across role types and identifying where these skills can fit within undergraduate physics curricula, this work supports educators in translating workforce needs into actionable instructional goals that can be addressed in the undergraduate curriculum.  

Future research can build on this work in several ways. Extending this work to include perspectives from a broader range of quantum company types could further refine and expand the experimental skills identified here. Additionally, investigating how quantum specific knowledge, skills, and abilities (KSAs) are defined across QISE careers and how they integrate with experimental skills would provide an even more comprehensive picture of how undergraduate courses prepare students for quantum industry roles. Finally, implementing the suggested instructional implications and investigating their impact on student learning and career trajectories in QISE would provide insight into how courses and programs support career preparation.

\begin{acknowledgments}
We would like to thank the quantum industry professionals who participated in our interviews. This work is based on work supported by the National Science Foundation under Grant Nos. PHY-2333073 and PHY-2333074. This work is also based on work supported by the Army Research Office and was accomplished under Award Number: W911NF-24-1-0132. 
\end{acknowledgments}

\onecolumngrid
\pagebreak

\appendix

\section{Categories of roles in the QISE industry} \label{appendix:categoriesofroles}

\begin{figure}[h!]
    \centering
    \includegraphics[width=0.9\linewidth]{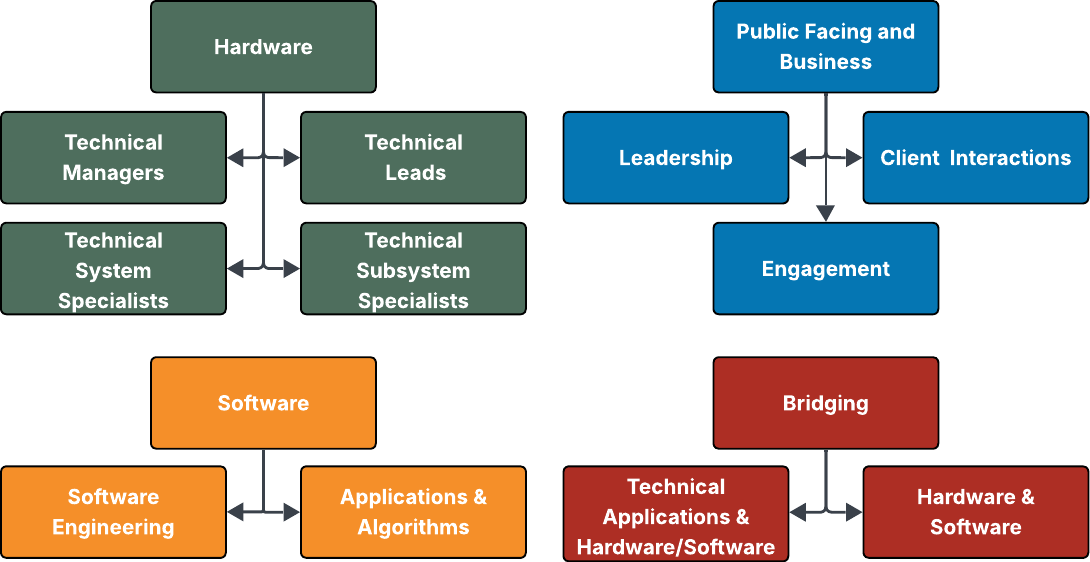}
    \caption{Major categories and their subcategories for roles in the QISE industry. Detailed description of the development of this categorization scheme is provided in \cite{pina2025categorization}.}
    \label{fig:categoriesQISEroles}
\end{figure}

\twocolumngrid
\pagebreak
\bibliography{apssamp}
\end{document}